\documentclass{aastex63}
\definecolor{green(pigment)}{rgb}{0.0, 0.65, 0.31}

\received{1 June 2020}
\revised{13 August 2020}
\accepted{14 August 2020}

\submitjournal{PSJ}

\graphicspath{{./}{figures/}}

\begin{document}

\title{Spatial variations of low mass negative ions in Titan's upper atmosphere}

\correspondingauthor{Teodora Mihailescu}
\email{teodora.mihailescu.19@ucl.ac.uk}

\author[0000-0001-8055-0472]{Teodora Mihailescu}
\affiliation{Blackett Laboratory \\
Imperial College London, UK}
\affiliation{Mullard Space Science Laboratory \\
University College London, UK}

\author[0000-0002-2015-4053]{Ravindra T. Desai}
\affiliation{Blackett Laboratory \\
Imperial College London, UK}

\author[0000-0001-9621-211X]{Oleg Shebanits}
\affiliation{Blackett Laboratory \\
Imperial College London, UK}

\author[0000-0002-5836-2827]{Richard Haythornthwaite}
\affiliation{Mullard Space Science Laboratory \\
University College London, UK}
\affiliation{The Centre for Planetary Science \\
University College London/Birkbeck, UK}

\author[0000-0002-2861-7999]{Anne Wellbrock}
\affiliation{Mullard Space Science Laboratory \\
University College London, UK}
\affiliation{The Centre for Planetary Science \\
University College London/Birkbeck, UK}

\author[0000-0002-6185-3125]{Andrew J. Coates}
\affiliation{Mullard Space Science Laboratory \\
University College London, UK}
\affiliation{The Centre for Planetary Science \\
University College London/Birkbeck, UK}

\author[0000-0003-4733-8319]{Jonathan P. Eastwood}
\affiliation{Blackett Laboratory \\
Imperial College London, UK}

\author[0000-0002-1978-1025]{J. Hunter Waite}
\affiliation{Southwest Research Institute \\
San Antonio, Texas, USA}

\begin{abstract}
 
Observations with Cassini's Electron Spectrometer discovered negative ions in Titan's ionosphere, at altitudes between 1400 and 950 km. Within the broad mass distribution extending up to several thousand amu, two distinct peaks were identified at 25.8-26.0 and 49.0-50.1 amu/q, corresponding to the carbon chain anions $CN^-$ and/or $C_2H^-$ for the first peak and $C_3N^-$ and/or $C_4H^-$ for the second peak. In this study we present the spatial distribution of these low mass negative ions from 28 Titan flybys with favourable observations between 26 October 2004 and 22 May 2012. We report a trend of lower densities on the night side and increased densities up to twice as high on the day side at small solar zenith angles. To further understand this trend, we compare the negative ion densities to the total electron density measured by Cassini's Langmuir Probe. We find the low mass negative ion density and the electron density to be proportional to each other on the dayside, but independent of each other on the night side. This indicates photochemical processes and is in agreement with the primary production route for the low mass negative ions being initiated by dissociative reactions with suprathermal electron populations produced by photoionisation. We also find the ratio of $CN^-/C_2H^-$ to $C_3N^-/C_4H^-$ highly constrained on the day-side, in agreement with this production channel, but notably displays large variations on the nightside. 

\end{abstract}


\section{Introduction}
Titan possesses a dense atmosphere composed of molecular nitrogen, methane and hydrogen \citep{Niemann2005, Coustenis2007} with a column density an order of magnitude larger than the Earth's and an abundance of complex organic chemistry \citep{Waite05, Vuitton2019}. A unique characteristic is the atmosphere containing opaque haze layers that block most visible light from the Sun, obscuring Titan's surface features. The Cassini mission studied the composition of the haze and mechanisms through which it could be produced and discovered large negatively charged ions, hereafter described as negative ions, at altitudes above 950 km \citep{Coates2007, Waite2007}. These surprising observations at such high altitudes indicated that these negative ions could be precursors of the aerosols forming the haze lower down.

Negative ions were initially detected with masses of up to 10,000 amu/q in Titan's upper atmosphere in the altitude region of 950km to 1400km using Cassini's Electron Spectrometer (ELS) \citep{Linder1998}, one of the sensors of the Cassini Plasma Spectrometer (CAPS) \citep{Young2004}. The discovery was based on data collected during 16 Titan flybys and was reported by \citet{Coates2007} and \citet{Waite2007}. Initial estimates showed that their density could reach $\sim$10\% of the negatively charged population, i.e. negative ions and electrons, at the lowest altitudes studied, around 950 km \citep{Coates2007}, with the remaining $\simeq90\%$ being electrons. The altitude, latitude and solar zenith angle (SZA) dependence of the negative ion mass was investigated by \citet{Coates2009} for 23 Titan encounters and the initial observations were summarised by \citet{Coates2010}. \citet{Coates2009} also provided the identification of larger masses up to 13,800 amu/q and revealed a preference for these larger masses at lower altitudes. They found that the highest detected ion mass increases with Titan latitude - with the largest masses detected near the north pole. 

The Radio and Plasma Wave Science (RPWS) Langmuir Probe (LP) \citep{Gurnett2004} was able to observe and confirm the presence of negative ions, as well as reveal that the negative ion densities are significant \citep{Agren2012, Shebanits2013, Shebanits2016}.  \citet{Agren2012} reported the detection of negative ion concentrations down to the lowest altitude of 880 km in the deep ionosphere (880–1000 km) during the T70 encounter (ELS did not detect negative ions during this flyby as the actuator was not pointing in the ram direction).  A study carried out by \citet{Wellbrock2013} on ELS data from 34 Titan flybys analysed the trends of peak densities for different negative ion mass ranges with altitude. The maximum altitude at which negative ions of a particular mass group were detected decreased with increasing mass revealing an ionosphere that becomes denser in negative ions as altitude decreases. In a similar way, \citet[][in preparation]{WellbrockAGU} investigate the densities of masses with SZA. For lower mass negative ions, they find a preference for higher densities on the dayside but find that heavy ions reach the highest densities on the night side. \citet{Shebanits2016} combined the LP observations with the ELS observations and estimated a net negative charge on the negative ions of $>$1. The authors also showed that the observed electron depletions exceeded 90\%, reaching as high as 96\%, which means that the negative ions are the main negative charge carriers, with densities almost matching those of the positive ions. 

The statistical identification of distinct species in the ELS data was provided by \citet{Desai2017}. This study modelled the instrument response function to negative ions and found the lightest negative ion detections were centred on 25.8-26.0 and 49.0-50.1 amu/q and correspond to the carbon chain anions, $CN^{-}$ and/or $C_{2}H^{-}$ for the first peak and $C_{3}N^{-}$ and/or $C_{4}H^{-}$ for the second peak. The logarithmic energy resolution of the ELS prohibited carrying out these precise identifications for the heavier species. Based on LP measurements, \citet{Shebanits2017} suggested that more heavy negative ions form at low solar EUV conditions. This is also consistent with the ELS measurements: more heavy negative ions at high latitudes during winter conditions (reduced EUV) reported by \citet{Wellbrock2019}, who studied seasonal effects on the heaviest negative ion species and focused on the T16 flyby that detected the heaviest negative ions of all observations. They found that the combination of high latitudes and winter conditions, i.e. where solar flux is restricted long-term, created the necessary environment for the heaviest negative ions to form. 
The larger negative ions might play a role in the formation of the organic macromolecules (aerosols) \citep{Waite2007,Coates2007, Lavvas2013} detected at lower altitudes via building processes, which then raises the question of how the spatial distribution and production mechanisms of low mass negative ions contribute to the formation of heavier species. 

In this paper, we report the spatial density variations of these low mass negative ions across Titan's upper atmosphere from data obtained by the ELS during 28 Titan flybys from 26 October 2004 to 22 May 2012. We compare these to the total electron densities measured by LP and discuss the results in relation to models of negative ion chemistry at Titan.

\begin{figure}
\includegraphics[width=17cm]{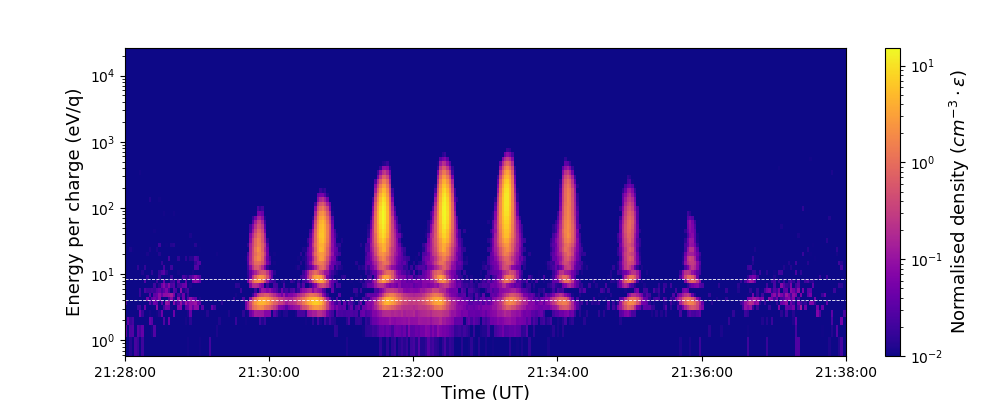}
\centering
\caption{Energy-time spectrogram produced by ELS during the T29 flyby, revealing negative ion signatures. The sharp vertical spikes correspond to times when the actuator in points in the ram direction, which indicates these are negative ion signatures rather than thermal electrons. The background electron counts are removed and the normalised negative ion densities calculated as outlined in the \textit{Method} section. The two individual peaks in density on the lower part of each of the signatures (marked with dashed white lines), correspond to the $CN^{-}/C_{2}H^{-}$ and $C_{3}N^{-}/C_{4}H^{-}$ signatures considered in this study.}
\label{neg_ion_signature}
\end{figure}

\section{Method}

The results presented in this study are based on data collected by CAPS-ELS during 28 Titan flybys where negative ions were detected. The ELS measures the raw count rate of negatively charged particles, which includes both electrons and negative ions. The ELS takes measurements across 8 anodes which cover $160^\circ \times 5^\circ$ of the 4$\pi$ space, see \citet[][Figure 7 therein]{Young2004} for a schematic of the ELS field-of-view. Two of these anodes typically point towards the ram direction (often anodes 4 and 5) and the others are tilted away from the ram direction. During Titan flybys, Cassini velocity varies between 5.7-6.3 km/s (supersonic). Negative ions are not entirely stationary, e.g. ion winds and thermal motion cause them to move within the atmosphere. However, since the spacecraft velocity is much higher, we can make the assumption that negative ions are stationary with respect to the atmosphere/ionosphere. This means that, when measured by ELS, they appear to only arrive from the ram direction, which allows for the energy-mass conversion. These anodes, however, also detect incoming electrons. Because electrons are thermalised, the electron count rate, i.e. the average of anodes not pointing in the spacecraft ram direction, can be subtracted as background noise. For example, anodes 2 and 7 are selected in this study in the majority of instances as they are tilted enough from the ram direction to not pick up the negative ions (anodes 3 and 6 sometimes do pick up negative ions as well) and also have a clear field of view (anodes 1 and 8 sometimes have an obstructed field of view). While a different choice of anodes might produce a different result, a consistent approach is used throughout. The negative ion counts were converted into density using the ion current approximation \citep{Waite05,Coates2007}:

\begin{equation}
\ n_{ni} = \frac{R_c}{v_{s/c}A\varepsilon},
\label{density}
\end{equation}
where $n_{ni}$ is the negative ion density, $R_c$ is the negative ion count rate, $v_{s/c}$ is the spacecraft velocity in Titan's reference frame, $A=0.33 cm^2$ \citep{Young2004} is the effective area of acceptance and $\varepsilon$ is the microchannel plate (MCP) ion detection efficiency. The instrument was originally designed to analyse electrons and the discovery of negative ions was unexpected, thus the response of the instrument was not calibrated for ions during laboratory tests. Hence, the efficiency can only be estimated using nominal ion efficiencies. Studies have shown that a value of $\varepsilon=0.05$ is appropriate for the largest species \citep{Fraser2002}. However, this value is higher for lower energies and can go up to $\varepsilon=0.5$ for the energy at which these low mass negative ions were detected \citep{Desai2018,Stephen2000}. Due to these uncertainties, all the relevant plots show the negative ion densities, normalised by the MCP efficiency, $n_{ni} \cdot \varepsilon$. Since the MCP efficiency uncertainty introduces an error that is mostly systematic, comparing the resulting densities within a mass group is accurate. An energy-time spectrogram corresponding to a set of negative ion signatures is shown in Figure \ref{neg_ion_signature}, with normalised density shown instead of counts. Each of the spikes corresponds to a negative ion signature detected when the ELS look direction scans across the ram direction, with the height being proportional to the mass of the heaviest ion detected.
The peak count rate during each scan across ram is selected although it should be noted that the negative ion densities could therefore be underestimated due to the instrument not precisely aligning with the incoming distribution \citep[][Chapter 4.2]{Desai_phd}. For this reason actuator fixed flybys T55-T59 have been omitted from this study.
The two individual peaks on the lower part of each negative ion signature correspond to two mass groups \citep{Wellbrock2013} which have been identified as primarily consisting of $CN^{-}/C_{2}H^{-}$ and $C_{3}N^{-}/C_{4}H^{-}$ \citep{Desai2017, Vuitton2009}. These two signatures are the focus of the present study.

\begin{figure}[h]
\includegraphics[width=18cm]{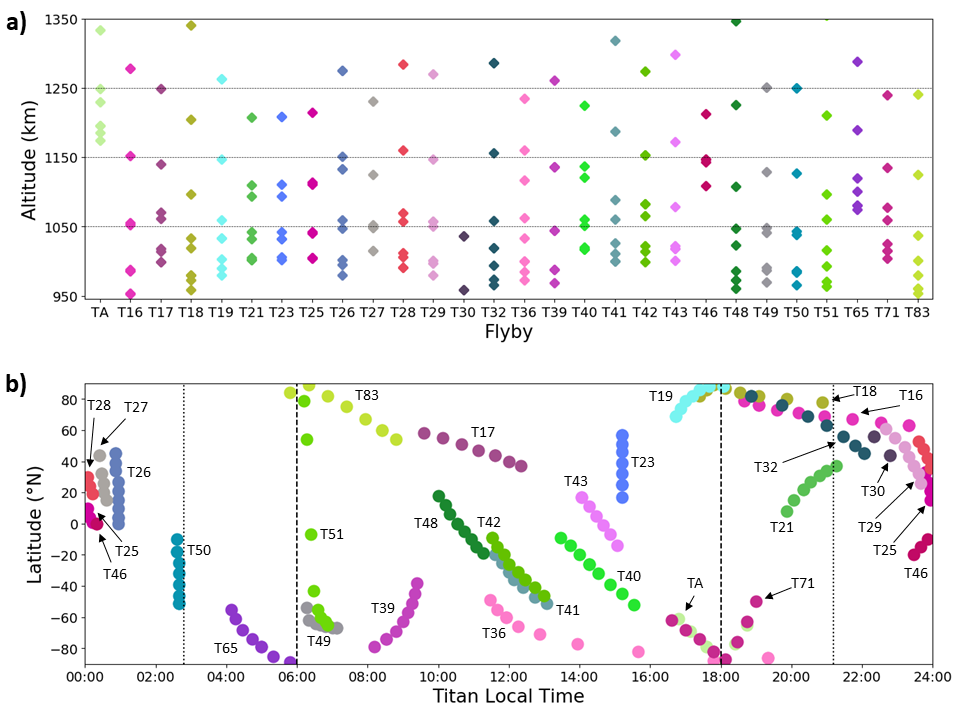}
\centering
\caption{Coverage of the data set used. The top plot shows the altitude distribution and the dashed black lines indicate the delimitation of the different altitude layers considered. Flybys TA-T51 took place when Titan was in southern summer and T65-T83 when Titan was in southern fall. The bottom plot shows the spatial coverage, the dashed lines indicate the separation between day side and night side, and the dotted lines show the delimitation of dusk/dawn (see \textit{Relation to production mechanisms} section).}
\label{titan_coverage}
\end{figure}

The timestamp, subsequently used to calculate the spacecraft position, corresponds to the highest density registered within the signature. Note that, within a negative ion signature, the density peak of the two mass groups do not necessarily align vertically (Figure \ref{neg_ion_signature}) and therefore have independent time stamps. We therefore conducted the studies for the two mass groups separately. For each negative ion signature, the density and spacecraft position were calculated to create the density distribution maps. A total of 225 data points are used and divided into 4 altitude ranges (Figure \ref{titan_coverage}) to generate global maps of the spatial distributions of the density within the atmosphere which varies significantly with altitude. The data set could be normalised by the local ionospheric density to account for the variations during the different flybys but the relatively large size of the altitude layers compared to the atmospheric scale height results in this being a reasonable approximation. Note that the spatial coverage of the ionosphere is limited and there is bias towards certain regions, i.e. high latitudes and at night and there are fewer measurements at higher altitudes.

Following this, we compare the results to the total electron densities measured by the LP. The ELS measures both electrons and negative ions, and the possibility of inter-anode cross-talk makes it difficult to differentiate between the two. The LP measures the plasma currents and can derive an precise electron density measurement although it cannot resolve individual ion species. A combination of the ELS and LP datasets is therefore beneficial. The ELS and LP measurements are are matched in time by linear interpolation (LP times to ELS times). The electron densities used in this study are derived from the ion current assuming quasineutrality, i.e. $n_e + Z_-n_- = n_+$, where $Z_-$ is charge number for negative ions/dust grains \citep{Shebanits2016, Shebanits2013}.

\section{Results and Analysis}
In this section, we first present the low mass negative ion density distribution results and compare them to the previously observed trends of the much heavier ions. We then compare to ionospheric electron density trends and discuss how the observed trends could be linked to negative ion production mechanisms. Finally, we discuss further external factors that could influence the density distributions.

\subsection{Density maps}
\begin{figure}
\includegraphics[width=18cm]{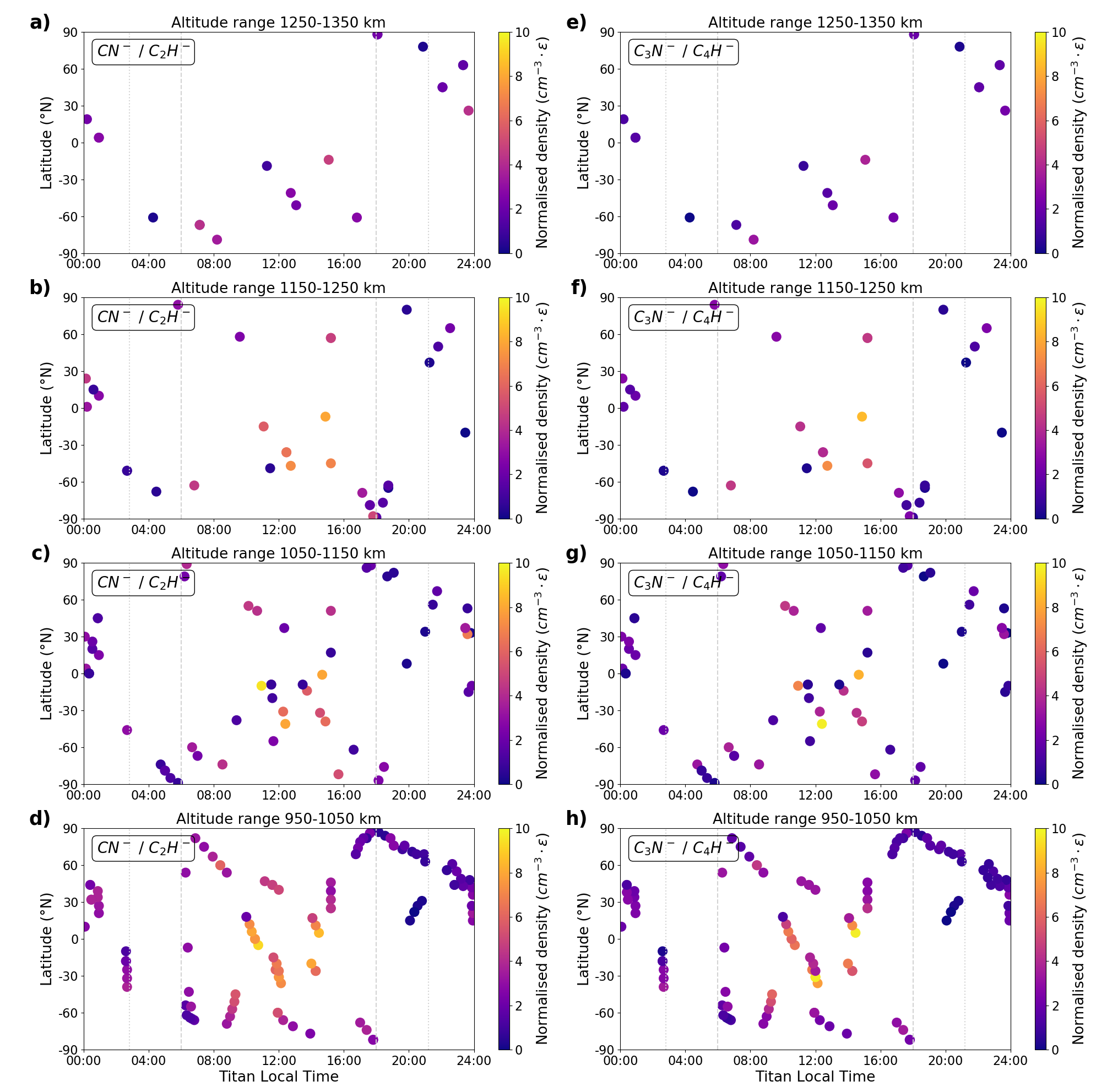}
\centering
\caption{Spatial density distribution of the lowest mass negative ion species, identified as $CN^-/C_2H^-$ (column of panels a-d on the left) and the second lowest mass negative ion species, identified as $C_3N^-/C_4H^-$ (column of panels e-h on the right), across Titan's upper atmosphere. Observations are divided into 4 altitude layers of 100km in thickness. Note that the x axis is Titan local time - this was chosen instead of longitude because Titan's atmosphere is photochemically induced. The dashed lines indicate the separation between day side and night side, and the dotted lines shows the delimitation of dusk/dawn (see \textit{Relation to production mechanisms} section).}
\label{species12}
\end{figure}

The results presented in Figure \ref{species12} show latitude vs local time maps of the low mass negative ions density. They were created by combining the density information shown in Figure \ref{neg_ion_signature} (and other flybys) and the location information shown in Figure \ref{titan_coverage}. The distributions shown in Figure \ref{species12}, panels a-d on the left column for $CN^-/C_2H^x-$ and e-h on the right column for $C_3N^-/C_4H^-$, show a high degree of similarity between the two negative ion mass groups and the trends are therefore discussed together.

The highest density values are concentrated between $-40^\circ$N - $20^\circ$N and 9:00-13:00 Titan local time (at midday). The lowest density values are registered in the northern hemisphere, at 16:00-24:00 Titan local time (mostly on the night side). Titan changed seasons in 2009, from southern summer/northern winter to southern fall/northern spring. The majority of the measurements, i.e. flybys TA (26 October 2004) to T51 (27 March 2009), were taken before the 2009 equinox, when Titan was in southern summer. Measurements from flybys T65 (12 January 2010) to T83 (22 May 2012) were taken during southern fall (see Figure \ref{titan_coverage}a). The overall trend is for lower densities on the night side and close to the poles and high densities close to midday and around the equator. This trend is most visible in the lowest altitude layer (950-1050 km), where the data coverage is highest (see Figure \ref{species12} panel d for $CN^-/C_2H^-$ and h for $C_3N^-/C_4H^-$). Going to higher altitudes (see Figure \ref{species12} panels a-c and e-g), this appears to still be true although less obvious. It is also unclear whether this SZA trend is predominantly a latitude or Titan local time effect due to the limited coverage of the ionosphere and bias towards certain regions, i.e. high latitudes at night.

The densities increase with decreasing altitude but the overall variations between altitude layers for both mass groups are not obvious here. \citet{Wellbrock2013} found that, for the first two mass groups, density decreases with altitude at a slower rate compared to the heavier mass groups. It is thus not surprising that we do not see obvious variations between altitude layers here. There are low and high negative ion densities at small SZA, especially at higher altitudes. This is an effect of binning large altitude ranges but could also be due to variabilities of Titan's ionosphere during different flybys. 

The low mass negative ions appear most abundant at small SZA. In contrast, the highest negative ion masses were preferentially found at high latitudes \citep{Coates2009} and in winter conditions \citep{Wellbrock2019}, while \citet{Desai2017} found a correlation between the low mass negative ions considered here and the larger negative ions, below the ionospheric peak. These results therefore raise the question of what production and loss mechanisms link the small and large species.

\subsection{Relation to production mechanisms}
\begin{figure}
\includegraphics[width=17cm]{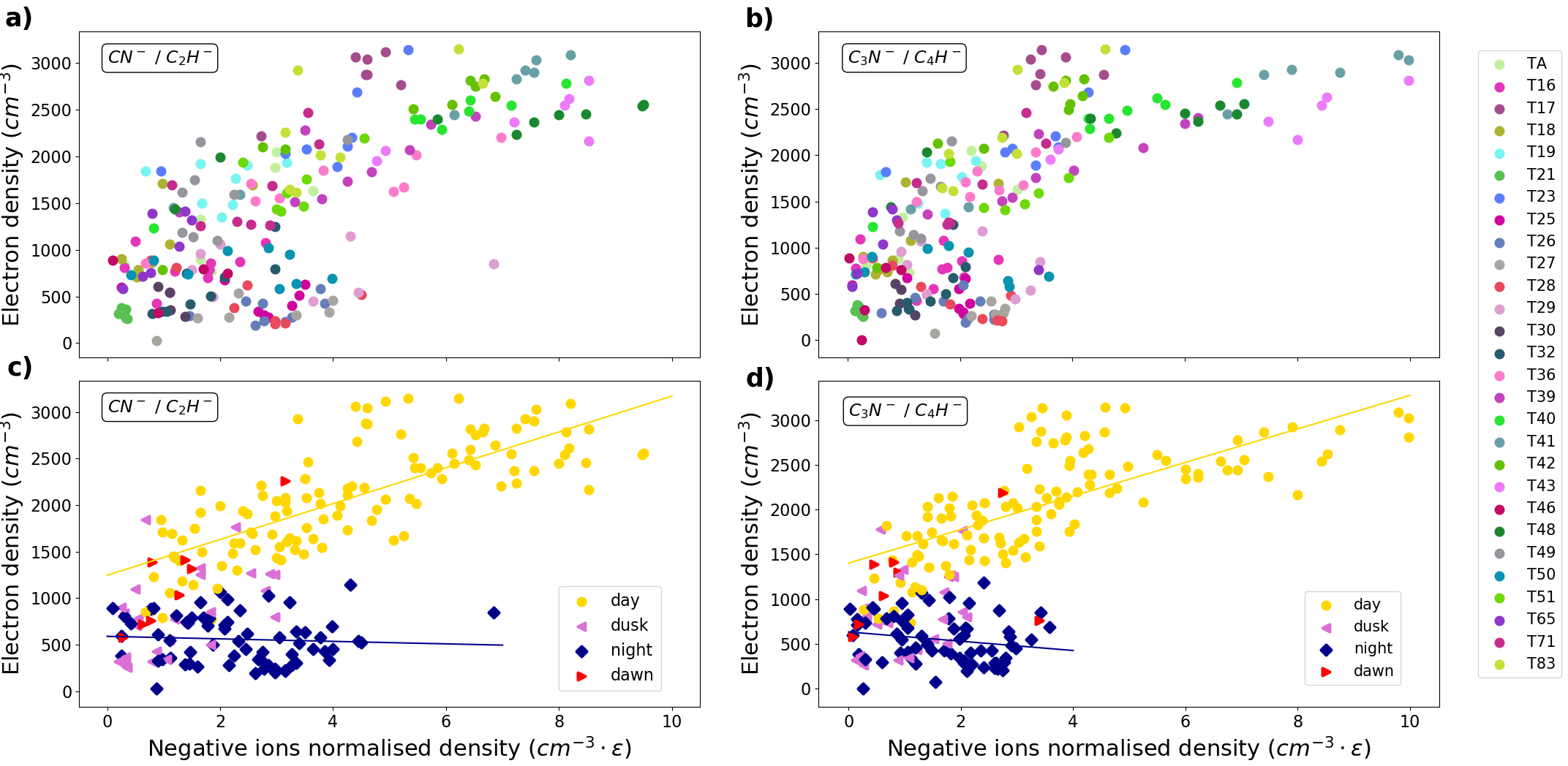}
\centering
\caption{Negative ion density as a function of electron density for the two species: $CN^-/C_2H^-$ (left) and $C_3N^-/C_4H^-$ (right). Colors showing the flyby during which the measurement was taken (top panel) and the location in Titan's atmosphere (bottom panel). }
\label{LPplot}
\end{figure}

To further understand the trends in Figure \ref{species12}, we compare our results to the total electron density distribution measured by LP in Figure \ref{LPplot}. All the data is included in the same figure which effectively removes inherent altitude density variations between the measurements.

In the upper two panels, two distinct distributions are present above and below an electron density of $\sim1000 cm^{-3}$. Above $\sim 1000 cm^{-3}$  the negative ion density increases with electron density while below there is no obvious dependence. The same data is then shown in the lower panels but differentiated according to Titan Local Time, divided into measurements taken on the day side and the night side.
Titan's atmosphere extends to an altitude of 1450 km which is almost half compared to its 2575km radius. This leaves large dawn and dusk regions at high altitudes where the solar photon flux is attenuated by the thick atmosphere. These dusk and dawn measurements are therefore marked separately in addition to the day and night measurements.

The densities appear proportional to each other on the day side, and independent of each other on the night side. The dusk and dawn `transition' regions fall in between the two distributions. The fits to these distributions shown in Figure \ref{LPplot} show that the negative ion density is approximately linearly proportional to the electron density on the day-side according to the relation:
\begin{equation}
    n_{ni} \varepsilon = (3.0 \pm 0.2) \times 10^{-3} n_e - (1.8 \pm 0.5) \hspace{5mm} [{cm^{-3}}{ \varepsilon}],
\end{equation}
for $CN^-/C_2H^-$, and:
\begin{equation}
    n_{ni} \varepsilon = (2.8 \pm 0.3)  \times 10^{-3} n_e - (2.1 \pm 0.5) \hspace{5mm} [{cm^{-3}}{ \varepsilon}],
\end{equation}
for $C_3N^-/C_4H^-$. There is however sufficient spread in the data which might be explained by further controlling factors. The fitted trend lines through the night-side negative ion densities, however, have very shallow gradients and the data spread is large, therefore indicating they are not related. 

Titan's thermal ionospheric plasma has been measured to be cold, $\sim$ 0.1 eV \citep{Crary2009}, although, on Titan's day-side, suprathermal electron populations are produced via photoionisation processes \citep{Haider1986,Galand2006,Coates2011}. \citet{Vuitton2009} produced the first negative ion chemistry model of Titan's ionosphere and identified that these suprathermal electrons are able to efficiently break the bonds within neutral species such as HCN which results in the dissociative reaction:
\begin{equation}
    \text{AB} + \text{e}_s^- \rightarrow \text{A}^- + \text{B}.
\label{dissociative_reaction}
\end{equation}
This was identified as the dominant production route for low mass negative ions  such as $CN^-$ and $C_2H^-$ \citep{Vuitton2009,Dobrijevic2016}, the main constituent species of the first mass group. The negative charge was then identified as being transferred via proton transfer reactions such as,
\begin{equation}
    \text{AH} + \text{B}^- \rightarrow \text{BH} + \text{A}^-,
\label{proton_transfer}
\end{equation}
onto heavier species and produce negative ions such as
$C_3N^-$ and $C_4H^-$, the main constituent species of the second mass group. \citet{Vuitton2009} also found that the main loss process is associative detachment with radicals ($H$ and $CH_3$). The results presented in Figure \ref{LPplot}, showing the negative ion and electron densities being proportional to each other on the day side, are in agreement with this proposed production channel. \citet{Dobrijevic2016}, with an updated photochemical model coupling negative and positive ions with neutral molecules, also found that the most abundant negative ions are $CN^-$ and $C_3N^-$, consistent with \citet{Vuitton2009}. \citet{Dobrijevic2016} used more recent cross section values for the dissociative electron attachment of $CH_4$ and $HCN$, however, which led them to conclude that the $H^-$ production rate is higher than $CN^-$, but also that $H^-$ is quickly transformed into $C_2H^-$ and $CN^-$ via a proton transfer with HCN. The calculated densities of the low mass negative ions in these chemical models \citep{Vuitton2009,Dobrijevic2016}, and also the recent modelling study of \citet{Mukundan2018} are, however, notably smaller than those measured by ELS. What is yet not clear from Figure \ref{LPplot} is whether the negative ions present on the night-side are locally produced via an additional production mechanism or are remnants of the higher densities produced on the day-side \citep{Agren2007, Cravens2009, Vigren2015}, possibly transported from day to night \citep{Muller-Wodarg2008, Cui2009}.

The spatial distribution of the low mass negative ions we present here is different from the distribution of the heavy negative ions, which were found preferentially where sunlight is attenuated or even absent \citep{Coates2009}. This is believed to be due to the combination of high latitude and winter conditions \citep{Wellbrock2019}. This appears in agreement with a large fraction of macromolecules being negative on the night-side but becoming neutral on the day side as they are not stable against photo detachment \citep{Bakes2002}. 

\begin{figure}
\includegraphics[width=9cm]{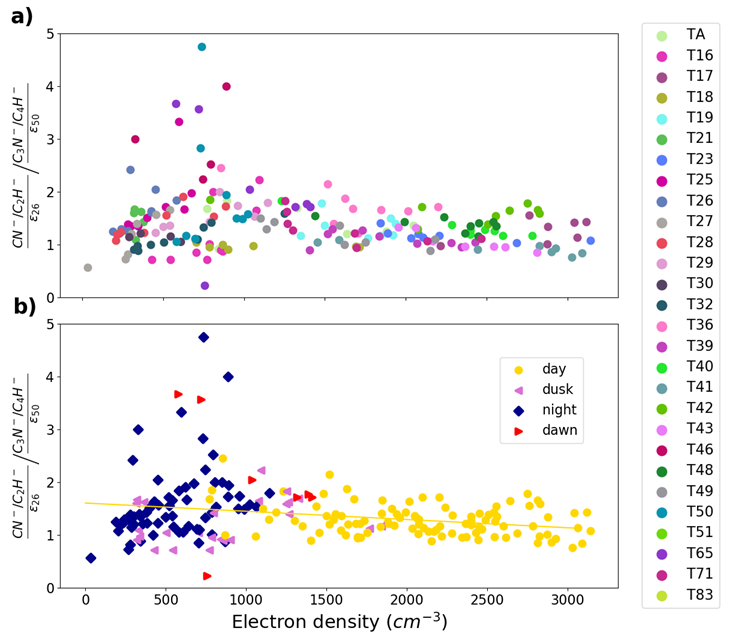}
\centering
\caption{Density ratio between $CN^-/C_2H^-$ (left) and $C_3N^-/C_4H^-$ plotted as a function of electron density. $CN^-/C_2H^-$ is normalised by the MCP efficiency of 26 amu, $\varepsilon_{26}$ and $C_3N^-/C_4H^-$ is normalised by the MCP efficiency of 50 amu.  $\varepsilon_{50}$. Colors showing the flyby during which the measurement was taken (top panel) and the location in Titan's atmosphere (bottom panel).}
\label{ratio}
\end{figure}

To quantitatively study the relationship between these low mass ions, Figure \ref{ratio} shows the density ratio of $CN^-/C_2H^-$ to $C_3N^-/C_4H^-$ as a function of the electron density. The results are normalised by the unknown MCP efficiencies. The MCP efficiency is, however, energy dependent \citep{Stephen2000,Fraser2002} with higher energy particles (more massive particles here as all negative ions arrive at the flyby velocity in the spacecraft frame) being detected more efficiently. The ratios presented here can therefore be treated as a reasonable lower bound.

The ratio appears approximately constant on dayside which indicates that the production and loss processes are highly correlated, as described in reactions (\ref{dissociative_reaction}) and (\ref{proton_transfer}). The constraints on this ratio across a wide range of electron densities also highlights the steady driving of Titan's ionospheric photochemistry. The ratios are, however, for nearly all datapoints, greater than 1. This indicates that the lower mass negative ions, $CN^-/C_2H^-$, are produced more efficiently or that $C_3N^-/C_4H^-$ are preferentially lost. There is also a slight gradient to the day-side ratios with larger ratios observed at higher altitudes. This is likely due to the lower masses possessing greater scale heights.

The night side ratios notably have a much greater spread with ratios extending up to 5 and to lower values too. The highest ratios are observed during T25, T46, T50 and T65 which all occur in the early morning, both within darkness and within the attenuated dawn atmosphere. There is, however, not enough coverage to constrain any latitudinal variations. As the $CN^-/C_2H^-$ and $C_3N^-/C_4H^-$ densities are highest during the day, where they primarily appear to be produced (Figure \ref{LPplot}), this trend in Titan local time suggests that $C_3N^-/C_4H^-$ is preferentially lost during the night or that $CN^-/C_2H^-$ is preferentially transported from day to night. Carbon chain anions can react with neutral molecules and polymerise into larger negative ions \citep{Deschenaux1999} although the rate coefficients for these reactions are unknown \citep{Vuitton2009}. \cite{Bakes2002} suggest building processes occur during the night when solar flux is absent and it is possible that $C_3N^-/C_4H^-$ preferentially contribute to the building processes \citep{Lavvas2013,Desai2017}. 

\subsection{Relation to other parameters}
The observed variation in density could also be influenced by a series of other factors. External factors include the magnetospheric electron flux \citep[e.g.][]{Galand2009} which is modulated by Titan's location with respect to the ambient plasma. Based on this, \citet{Rymer2009} classified the Titan flybys into four distinct groups: plasma-sheet (high energy and density electrons), lobe-like (high energy but lower density electrons), magnetosheath (outside the magnetopause, within the shocked solar wind - first observation of this reported by \citet{Bertucci2008}) and bimodal (two distinct electron populations). They provide typical electron properties for these regimes which might have an effect on the measured negative ion densities. EUV variability was also studied using the LP measurements of electrons by \citet{Edberg2013} and of ions by \citet{Shebanits2017}, and, using the ELS measurements of negative ions, by \citet{Wellbrock2019}. A comparison between negative ion densities and specific parts of the electron spectra might also offer insight into which electrons are initiating the negative ion production and loss reactions. The relation to ion and neutral spatial distributions might also provide information into coupled ion-neutral chemical processes. The finite number of measurements across this multi-dimensional parameter space prohibits the simple analysis of these variables which are left for a further study.

\citet{Dubois2019} also reported the presence a number of other species such as CNN$^-$ and CHNN$^-$ in laboratory experiments of aerosol growth and a corresponding peak in the ELS spectra. The identification and distributions of further such negative ion species in the ELS data, and also produced in laboratory experiments of Titan's atmospheric chemistry \citep[e.g.][]{Hoerst2012,Trainer2013,Dubois2019}, could also help unravel the various controlling factors involved in negative ion production and loss processes. 

\section{Summary}

We studied the spatial density variation of the two lightest negative ion mass groups detected in Titan's upper atmosphere, previously identified as primarily consisting of $CN^-/C_2H^-$ and $C_3N^-/C_4H^-$ \citep{Desai2017}. The study revealed similar trends for the two mass groups: increased density at small SZA, i.e. on the day side and around the equator. This trend was observed mainly in the 950-1050 km altitude region, where there were more data points. The trends seemed to carry on at higher altitudes although they are less obvious. The high negative ion densities at midday appeared to suggest a dependency on photo-ionisation. Comparison with LP measurements revealed a linear dependence of negative ion density to electron density on the day side and no dependence on the night side, for both species. This trend indicates that the low mass negative ions are produced as a result of photochemical processes and provides evidence for the dominant production pathways of dissociative reactions involving suprathermal electron populations, as outlined by \citet{Vuitton2009}, which only exist on Titan's day-side. The spatial distributions of $CN^-/C_2H^-$ and $C_3N^-/C_4H^-$ appeared similar and the ratio of the two appeared highly constrained on the day-side therefore further supporting similar and inter-related production pathways. On the nightside, however, the ratio displayed large variations extending up to much greater than 1. This is suggested to result from preferential loss or day-to-night transport. In contrast to the spatial distributions of the low mass negative ions, the highest mass negative ions have been preferentially observed at high latitudes \citep{Coates2009}. This raises interesting questions about how the production and loss processes of the low mass negative ions are related to those of the heavier negative ions. 

\acknowledgements
The authors thank two reviewers for their careful comments which improved this manuscript.  R.T.D. acknowledges funding from the NERC grant NE/P017347/1 (Rad-Sat). O.S. acknowledges funding by the Royal Society grant RP$\backslash$EA$\backslash$180014. A.J.C. and A.W. acknowledge the support of STFC grant ST/S000240/1. This work has benefited from discussions within International Space Science Institute (ISSI) International Team 437. The Cassini dataset presented in this study is publicly available from the NASA Planetary Data System archive.

\newpage

\begin{thebibliography}{}
\expandafter\ifx\csname natexlab\endcsname\relax\def\natexlab#1{#1}\fi
\providecommand{\url}[1]{\href{#1}{#1}}
\providecommand{\dodoi}[1]{doi:~\href{http://doi.org/#1}{\nolinkurl{#1}}}
\providecommand{\doeprint}[1]{\href{http://ascl.net/#1}{\nolinkurl{http://ascl.net/#1}}}
\providecommand{\doarXiv}[1]{\href{https://arxiv.org/abs/#1}{\nolinkurl{https://arxiv.org/abs/#1}}}

\bibitem[{Agren {et~al.}(2007)Agren, Wahlund, Modolo, Lummerzheim, Galand,
  Müller-Wodarg, Canu, Cravens, Jr., Coates, Lewis, Young, Bertucci, , \&
  Dougherty}]{Agren2007}
Agren, K., Wahlund, J.-E., Modolo, R., {et~al.} 2007, Annales Geophysicae, 25,
  2359–2369

\bibitem[{Bakes {et~al.}(2002)Bakes, Mckay, \& Bauschlicher}]{Bakes2002}
Bakes, E., Mckay, C.~P., \& Bauschlicher, C.~W. 2002, Icarus, 157, 464

\bibitem[{Bertucci {et~al.}(2008)Bertucci, Achilleos, Dougherty, Modolo,
  Coates, Szego, Masters, Ma, Neubauer, Garnier, Wahlund, \&
  Young}]{Bertucci2008}
Bertucci, C., Achilleos, N., Dougherty, M.~K., {et~al.} 2008, Science, 321,
  1475.
\newblock \url{http://www.jstor.org/stable/20144798}

\bibitem[{Coates {et~al.}(2009)Coates, Wellbrock, Lewis, Jones, Young, Crary,
  \& Waite}]{Coates2009}
Coates, A., Wellbrock, A., Lewis, G., {et~al.} 2009, Planetary and Space
  Science, 57, 1866 , \dodoi{https://doi.org/10.1016/j.pss.2009.05.009}

\bibitem[{Coates {et~al.}(2007)Coates, Crary, Lewis, Young, Waite~Jr., \&
  Sittler~Jr.}]{Coates2007}
Coates, A.~J., Crary, F.~J., Lewis, G.~R., {et~al.} 2007, Geophysical Research
  Letters, 34, \dodoi{10.1029/2007GL030978}

\bibitem[{Coates {et~al.}(2010)Coates, Wellbrock, Lewis, Jones, Young, Crary,
  Waite, Johnson, Hill, \& Sittler~Jr.}]{Coates2010}
Coates, A.~J., Wellbrock, A., Lewis, G.~R., {et~al.} 2010, Faraday Discussions,
  147, 293

\bibitem[{{Coates} {et~al.}(2011){Coates}, {Tsang}, {Wellbrock}, {Frahm},
  {Winningham}, {Barabash}, {Lundin}, {Young}, \& {Crary}}]{Coates2011}
{Coates}, A.~J., {Tsang}, S.~M.~E., {Wellbrock}, A., {et~al.} 2011, \planss,
  59, 1019, \dodoi{10.1016/j.pss.2010.07.016}

\bibitem[{Coustenis {et~al.}(2007)Coustenis, Achterberg, Conrath, Jennings,
  Marten, Gautier, Nixon, Flasar, Teanby, Bezard, Samuelson, Carlson, Lellouch,
  Bjoraker, Romani, Taylor, Irwin, Fouchet, Hubert, Orton, Kunde, Vinatier,
  Mondellini, Abbas, \& Courtin}]{Coustenis2007}
Coustenis, A., Achterberg, R., Conrath, B., {et~al.} 2007, Icarus, 189, 35

\bibitem[{{Crary} {et~al.}(2009){Crary}, {Magee}, {Mandt}, {Waite}, {Westlake},
  \& {Young}}]{Crary2009}
{Crary}, F.~J., {Magee}, B.~A., {Mandt}, K., {et~al.} 2009, \planss, 57, 1847,
  \dodoi{10.1016/j.pss.2009.09.006}

\bibitem[{Cravens {et~al.}(2009)Cravens, Robertson, Waite, Yelle, Vuitton,
  Coates, Wahlund, Agren, Richard, De~La~Haye, Wellbrock, \&
  Neubauer}]{Cravens2009}
Cravens, T., Robertson, I., Waite, J., {et~al.} 2009, Icarus, 199, 174

\bibitem[{Cui {et~al.}(2010)Cui, Galand, Yelle, Wahlund, Ågren, Waite~Jr., \&
  Dougherty}]{Cui2009}
Cui, J., Galand, M., Yelle, R.~V., {et~al.} 2010, Journal of Geophysical
  Research: Space Physics, 115

\bibitem[{{Desai}(2018)}]{Desai_phd}
{Desai}, R.~T. 2018, {Negative Ions in Outer Solar System Plasmas}, PhD thesis,
  University College London, UK.
  https://discovery.ucl.ac.uk/id/eprint/10049925/

\bibitem[{Desai {et~al.}(2017)Desai, Coates, Wellbrock, Vuitton, Crary,
  Gonz{\'{a}}lez-Caniulef, Shebanits, Jones, Lewis, Waite, Cordiner, Taylor,
  Kataria, Wahlund, Edberg, \& Sittler}]{Desai2017}
Desai, R.~T., Coates, A.~J., Wellbrock, A., {et~al.} 2017, The Astrophysical
  Journal, 844, L18, \dodoi{10.3847/2041-8213/aa7851}

\bibitem[{{Desai} {et~al.}(2018){Desai}, {Taylor}, {Regoli}, {Coates},
  {Nordheim}, {Cordiner}, {Teolis}, {Thomsen}, {Johnson}, {Jones}, {Cowee}, \&
  {Waite}}]{Desai2018}
{Desai}, R.~T., {Taylor}, S.~A., {Regoli}, L.~H., {et~al.} 2018, \grl, 45,
  1704, \dodoi{10.1002/2017GL076588}

\bibitem[{Deschenauxdag {et~al.}(1999)Deschenauxdag, Affolterdag, Magnidag,
  Hollensteindag, \& P.}]{Deschenaux1999}
Deschenauxdag, C., Affolterdag, A., Magnidag, D., Hollensteindag, C., \& P., F.
  1999, Journal of Physics D: Applied Physics, 32

\bibitem[{Dobrijevic {et~al.}(2016)Dobrijevic, Loison, Hickson, \&
  Gronoff}]{Dobrijevic2016}
Dobrijevic, M., Loison, J., Hickson, K., \& Gronoff, G. 2016, Icarus, 268, 313

\bibitem[{{Dubois} {et~al.}(2019){Dubois}, {Carrasco}, {Bourgalais}, {Vettier},
  {Desai}, {Wellbrock}, \& {Coates}}]{Dubois2019}
{Dubois}, D., {Carrasco}, N., {Bourgalais}, J., {et~al.} 2019, \apjl, 872, L31,
  \dodoi{10.3847/2041-8213/ab05e5}

\bibitem[{Edberg {et~al.}(2013)Edberg, Andrews, Shebanits, Ågren, Wahlund,
  Opgenoorth, Cravens, \& Girazian}]{Edberg2013}
Edberg, N. J.~T., Andrews, D.~J., Shebanits, O., {et~al.} 2013, Journal of
  Geophysical Research: Space Physics, 118, 5255

\bibitem[{Fraser(2002)}]{Fraser2002}
Fraser, G. 2002, International Journal of Mass Spectrometry, 215, 13

\bibitem[{Galand {et~al.}(2009)Galand, Moore, Charnay, Mueller‐Wodarg, \&
  Mendillo}]{Galand2009}
Galand, M., Moore, L., Charnay, B., Mueller‐Wodarg, I., \& Mendillo, M. 2009,
  Journal of Geophysical Research: Space Physics, 114, n/a

\bibitem[{Galand {et~al.}(2006)Galand, Yelle, Coates, Backes, \&
  Wahlund}]{Galand2006}
Galand, M., Yelle, R., Coates, A., Backes, H., \& Wahlund, J. 2006, Geophysical
  Research Letters , 33 (21) , Article L21101. (2006)

\bibitem[{Gurnett {et~al.}(2004)Gurnett, Kurth, Kirchner, Hospodarsky,
  Averkamp, Zarka, Lecacheux, Manning, Roux, Canu, Cornilleau-Wehrlin,
  Galopeau, Meyer, Boström, Gustafsson, Wahlund, Åhlen, Rucker, Ladreiter,
  Macher, Woolliscroft, Alleyne, Kaiser, Desch, Farrell, Harvey, Louarn,
  Kellogg, Goetz, \& Pedersen}]{Gurnett2004}
Gurnett, D., Kurth, W., Kirchner, D., {et~al.} 2004, Space Science Reviews,
  114, 395

\bibitem[{{Haider}(1986)}]{Haider1986}
{Haider}, S.~A. 1986, \jgr, 91, 8998, \dodoi{10.1029/JA091iA08p08998}

\bibitem[{{H{\"o}rst} {et~al.}(2012){H{\"o}rst}, {Yelle}, {Buch}, {Carrasco},
  {Cernogora}, {Dutuit}, {Quirico}, {Sciamma-O'Brien}, {Smith}, {Somogyi},
  {Szopa}, {Thissen}, \& {Vuitton}}]{Hoerst2012}
{H{\"o}rst}, S.~M., {Yelle}, R.~V., {Buch}, A., {et~al.} 2012, Astrobiology,
  12, 809, \dodoi{10.1089/ast.2011.0623}

\bibitem[{Lavvas {et~al.}(2013)Lavvas, Yelle, Koskinen, Bazin, Vuitton, Vigren,
  Galand, Wellbrock, Coates, Wahlund, Crary, \& Snowden}]{Lavvas2013}
Lavvas, P., Yelle, R.~V., Koskinen, T., {et~al.} 2013, Proceedings of the
  National Academy of Sciences, 110, 2729

\bibitem[{Linder {et~al.}(1998)Linder, Coates, Woodliffe, Alsop, Johnstone,
  Grande, Preece, Narheim, \& Young}]{Linder1998}
Linder, D., Coates, A., Woodliffe, R., {et~al.} 1998, in Geophysical Monograph
  Series, Vol. 102 (Blackwell Publishing Ltd), 257--262

\bibitem[{Mukundan \& Bhardwaj(2018)}]{Mukundan2018}
Mukundan, V., \& Bhardwaj, A. 2018, Icarus, 299, 222

\bibitem[{Müller-Wodarg {et~al.}(2008)Müller-Wodarg, Yelle, Cui, \&
  Waite}]{Muller-Wodarg2008}
Müller-Wodarg, I. C.~F., Yelle, R.~V., Cui, J., \& Waite, J.~H. 2008, Journal
  of Geophysical Research: Planets, 113

\bibitem[{Niemann {et~al.}(2005)Niemann, Atreya, Bauer, \&
  et~al.}]{Niemann2005}
Niemann, H., Atreya, S., Bauer, S., \& et~al. 2005, Nature, 438, 779–784

\bibitem[{{{\r{A}}gren} {et~al.}(2012){{\r{A}}gren}, {Edberg}, \&
  {Wahlund}}]{Agren2012}
{{\r{A}}gren}, K., {Edberg}, N.~J.~T., \& {Wahlund}, J.~E. 2012, \grl, 39,
  L10201

\bibitem[{Rymer {et~al.}(2009)Rymer, Smith, Wellbrock, Coates, \&
  Young}]{Rymer2009}
Rymer, A., Smith, H., Wellbrock, A., Coates, A., \& Young, D. 2009, Geophysical
  Research Letters , 36 , Article L15109. (2009)

\bibitem[{Shebanits {et~al.}(2017)Shebanits, Vigren, Wahlund, Holmberg,
  Morooka, Edberg, Mandt, \& Waite}]{Shebanits2017}
Shebanits, O., Vigren, E., Wahlund, J.-E., {et~al.} 2017, Journal of
  Geophysical Research: Space Physics, 122, 7491

\bibitem[{{Shebanits} {et~al.}(2013){Shebanits}, {Wahlund}, {Mandt},
  {{\r{A}}gren}, {Edberg}, \& {Waite}}]{Shebanits2013}
{Shebanits}, O., {Wahlund}, J.~E., {Mandt}, K., {et~al.} 2013, \planss, 84, 153

\bibitem[{{Shebanits} {et~al.}(2016){Shebanits}, {Wahlund}, {Edberg}, {Crary},
  {Wellbrock}, {Andrews}, {Vigren}, {Desai}, {Coates}, {Mandt}, \&
  {Waite}}]{Shebanits2016}
{Shebanits}, O., {Wahlund}, J.~E., {Edberg}, N.~J.~T., {et~al.} 2016, Journal
  of Geophysical Research (Space Physics), 121, 10,075

\bibitem[{Stephen \& Peko(2000)}]{Stephen2000}
Stephen, T., \& Peko, B. 2000, Review Of Scientific Instruments, 71, 1355

\bibitem[{{Trainer} {et~al.}(2013){Trainer}, {Sebree}, {Yoon}, \&
  {Tolbert}}]{Trainer2013}
{Trainer}, M.~G., {Sebree}, J.~A., {Yoon}, Y.~H., \& {Tolbert}, M.~A. 2013,
  \apjl, 766, L4, \dodoi{10.1088/2041-8205/766/1/L4}

\bibitem[{Vigren {et~al.}(2015)Vigren, Galand, Yelle, Wellbrock, Coates,
  Snowden, Cui, Lavvas, Edberg, Shebanits, Wahlund, Vuitton, \&
  Mandt}]{Vigren2015}
Vigren, E., Galand, M., Yelle, R., {et~al.} 2015, Icarus, 248, 539

\bibitem[{Vuitton {et~al.}(2009)Vuitton, Lavvas, Yelle, Galand, Wellbrock,
  Lewis, Coates, \& Wahlund}]{Vuitton2009}
Vuitton, V., Lavvas, P., Yelle, R., {et~al.} 2009, Planetary and Space Science,
  57, 1558

\bibitem[{Vuitton {et~al.}(2019)Vuitton, Yelle, Klippenstein, Hörst, \&
  Lavvas}]{Vuitton2019}
Vuitton, V., Yelle, R., Klippenstein, S., Hörst, S., \& Lavvas, P. 2019,
  Icarus, 324, 120 , \dodoi{https://doi.org/10.1016/j.icarus.2018.06.013}

\bibitem[{Waite {et~al.}(2007)Waite, Young, Cravens, Coates, Crary, Magee, \&
  Westlake}]{Waite2007}
Waite, J.~H., Young, D.~T., Cravens, T.~E., {et~al.} 2007, Science (New York,
  N.Y.), 316, 870

\bibitem[{{Waite} {et~al.}(2005){Waite}, {Niemann}, {Yelle}, {Kasprzak},
  {Cravens}, {Luhmann}, {McNutt}, {Ip}, {Gell}, {De La Haye},
  {M{\"u}ller-Wordag}, {Magee}, {Borggren}, {Ledvina}, {Fletcher}, {Walter},
  {Miller}, {Scherer}, {Thorpe}, {Xu}, {Block}, \& {Arnett}}]{Waite05}
{Waite}, J.~H., {Niemann}, H., {Yelle}, R.~V., {et~al.} 2005, Science, 308,
  982, \dodoi{10.1126/science.1110652}

\bibitem[{Wellbrock {et~al.}(2019)Wellbrock, Coates, Jones, Vuitton, Lavvas,
  Desai, \& Waite}]{Wellbrock2019}
Wellbrock, A., Coates, A., Jones, G., {et~al.} 2019, Monthly Notices of the
  Royal Astronomical Society (2019), 490, 2254–2261,
  \dodoi{10.1093/mnras/stz2655}

\bibitem[{Wellbrock {et~al.}(2012)Wellbrock, Coates, Jones, Arridge, Lewis,
  Sittler, \& Young}]{WellbrockAGU}
Wellbrock, A., Coates, A.~J., Jones, G.~H., {et~al.} 2012, American Geophysical
  Union, Fall Meeting 2012

\bibitem[{Wellbrock {et~al.}(2013)Wellbrock, Coates, Jones, Lewis, \&
  Waite}]{Wellbrock2013}
Wellbrock, A., Coates, A.~J., Jones, G.~H., Lewis, G.~R., \& Waite, J.~H. 2013,
  Geophysical Research Letters, 40, 4481, \dodoi{10.1002/grl.50751}

\bibitem[{Young {et~al.}(2004)Young, Berthelier, Blanc, Burch, Coates,
  Goldstein, Grande, Hill, Johnson, Kelha, Mccomas, Sittler, Svenes, Szegö,
  Tanskanen, Ahola, Anderson, Bakshi, Baragiola, Barraclough, Black, Bolton,
  Booker, Bowman, Casey, Crary, Delapp, Dirks, Eaker, Funsten, Furman, Gosling,
  Hannula, Holmlund, Huomo, Illiano, Jensen, Johnson, Linder, Luntama, Maurice,
  Mccabe, Mursula, Narheim, Nordholt, Preece, Rudzki, Ruitberg, Smith, Szalai,
  Thomsen, Viherkanto, Vilppola, Vollmer, Wahl, Wüest, Ylikorpi, \&
  Zinsmeyer}]{Young2004}
Young, D., Berthelier, J., Blanc, M., {et~al.} 2004, Space Science Reviews,
  114, 1

\end{thebibliography}

 \newcommand{\noop}[1]{}

\bibliographystyle{aasjournal}

\end{document}